\documentclass[twocolumn,showpacs,amssymb,nobibnotes,aps,floats,psfig,prb]{revtex4-1}
\usepackage{textcomp,amssymb,graphicx,epsf}
\usepackage{hyperref}
\usepackage{amsbsy}
\usepackage{amsmath}
\usepackage{amssymb}
\usepackage[dvips]{color}
\usepackage{graphicx,epsfig}
\usepackage{epsfig}
\usepackage{float}
\usepackage{multirow}
\date{}

\def \be{\begin{eqnarray}}
\def \ee{\end{eqnarray}}

\def \sd{\sum\limits}

\begin{document}
\title{An analysis of the $t_2-V$
model}
\author{A. Ghosh${^{1}}$}
\author{S. Yarlagadda${^{1,2}}$}
\affiliation{${^1}$ TCMP Div., Saha Institute of Nuclear Physics,
Kolkata, India}
\affiliation{${^2}$Cavendish Lab, Univ. of Cambridge, Cambridge, UK}

\date{\today}
\begin{abstract}

We study a  model (i.e., the $t_2-V$ model involving {\em next-nearest-neighbor hopping} and nearest-neighbor repulsion) in one-dimension 
that generically depicts the dominant transport mechanism
in cooperative strong electron-phonon interaction systems.
Using 
analytic  and numerical approaches,  hard-core-bosons 
 are shown to typically undergo a 
striking discontinuous transition from a superfluid to a supersolid. 
Topological inequivalence of  rings with even and odd number of sites is manifested  through  
  observable 
 differences  (in structure factor peaks) at the transition.
Connections are also identified between the $t_2-V$ model  and other topologically interesting models.
\end{abstract}
\maketitle
\section{Introduction}
The last few decades have witnessed numerous
 studies to fathom the tapestry of
exotic phenomena (such as 
long-range 
orderings)   and 
interesting functionalities (such as colossal magnetoresistance, multiferroicity, superconductivity, etc.) \cite{hwang}
 in bulk  
transition metal oxides (such as the manganites, cuprates, etc.) and their interfaces.
Of considerable interest is the coexistence of diagonal long-range orders
[such as the charge-density-wave (CDW), spin-density-wave (SDW) and orbital-density-wave (ODW)
in manganites \cite{hotta}]; also of immense focus is the coexistence and competition between long-range orders that are
diagonal (i.e., CDW or SDW) and off-diagonal  (i.e., superconductivity or superfluidity)
such as those reported in bismuthates \cite{blanton}, cuprates \cite{abbamonte}, etc.

To model the emergent ordering and functionality in these complex metal oxides
(and guide material synthesis), one needs,
as building blocks,
effective Hamiltonians for various interactions.
Except for the cooperative electron-phonon interaction (EPI),
effective Hamiltonians, that reasonably mimic the physics,
 have been derived for all other interactions. For instance,
double exchange model approximates infinite Hund's coupling, Gutzwiller
approximation or dynamical mean-field theory model
 Hubbard on-site Coulombic interaction, superexchange
describes localized spin interaction at strong on-site repulsion, etc. 
Many oxides such as 
cuprates \cite{photoem1,photoem3,damascelli},
 manganites \cite{lanzara2,pbl,boothroyd}, and bismuthates \cite{tvr1}
 indicate cooperative strong EPI.

Although definite progress has been made long ago in numerically treating
EPI systems \cite{hotta}, only recently has the effective Hamiltonian 
been derived for the cooperative  EPI quantum systems in one-dimension;
it has been demonstrated analytically that introducing  cooperative effects in the strong EPI
limit changes the dominant transport mechanism  from
one of nearest-neighbor (NN) hopping to that of next-nearest-neighbor (NNN) hopping \cite{rpsy}.
Additional NN particle repulsion 
 (due to incompatibility of distortions produced by cooperative EPI effects) leads to the $t_2-V$ model as the effective model.

The purpose of the present paper is to study the $t_2-V$ model and elucidate the consequences
of the atypical  dominant (NNN) transport mechanism in cooperative strong EPI systems. 
We demonstrate that the $t_2-V$ model, upon tuning repulsion, 
displays a dramatic discontinuous transition from a superfluid 
 to either a CDW or a supersolid 
wherein the
superfluid and the CDW coexist instead  of compete.
Green's function analysis yields
exact critical repulsion values $V_c$ in the two limiting
cases; we find $V_c/t_2=4 $ for the two hard-core-boson (HCB) case and $V_c/t_2= 2\sqrt{2}$ for
the half-filled system. 
Using finite size scaling analysis,
 we also obtain $V_c$ values numerically at intermediate fillings.
The symmetry difference between rings with odd number of sites (o-rings) and rings 
with even number of sites (e-rings) 
is revealed through the ratio of their structure factor peaks at transition
 being an irrational number $4/\pi^2$.

{The paper is organized as follows. In Sec. \ref{sec2},
 we present formulae for the structure factor and superfluid fraction
(for both $V=0$ and $V>V_c$) and  obtain  numerical results for the $t_2-V$ model in e-rings.
 Next, in Secs. \ref{sec3} and 
\ref{sec4}, we obtain the exact instability conditions analytically for 
 e-rings at half-filling and for e-rings with two HCBs, 
respectively. Then for o-rings, in  Sec. \ref{sec5},
we study numerically the structure factor and superfluid fraction and also 
 derive  relevant formulae for them. 
 A comparison is made  between e-rings and o-rings in Sec. \ref{sec6}. In Sec. \ref{sec7}, we explain
why the Bose-Einstein condensate fraction is zero for our $t_2-V$ model; in Sec. \ref{sec8} 
we discuss the connection between our $t_2-V$ model and other models.
 Finally, we discuss our results in Sec. \ref{sec9} and present our conclusions
in Sec. \ref{sec10}.}

\section {Numerical study of the $t_2-V$ model for \lowercase{e-rings}} \label{sec2}
We begin by identifying the Hamiltonian of the $t_2-V$ model for HCBs.
\begin{eqnarray}
H_{t_2V} \equiv &&  
 - t_2 \sum_{j=1}^{N_s} (b^{\dagger}_{j-1} b_{j+1} + {\rm H.c.} ) 
+ V \sum_{j=1}^{N_s} n_{j} n_{j+1} ,
\label{eq:t2V}
 \end{eqnarray} 
where $b_j$ is the destruction operator for a HCB, $V \ge 0$,  $n_j =b^{\dagger}_j b_j$, and $N_s$ 
is the total number
of sites.
We assume periodic boundary conditions and first study numerically 
(using modified Lanczos algorithm \cite{dagotto})
the quantum phase transition (QPT) in the $t_2-V$ model.
 We will characterize the 
 transition
through the structure factor and the superfluid density.
   The structure factor
is given by
$ S(k)=\sum_{l=1}^{N_s}e^{ikl}W(l) $
where $W(l)$ is the two-point correlation function for density fluctuations of HCBs at
 a  distance $l$ apart (when lattice constant is set to unity): 
$ W(l)=\frac{4}{N_s}\sum_{j=1}^{N_s}\left[\langle n_jn_{j+l}\rangle-\langle n_j\rangle\langle n_{j+l}\rangle\right] $.
The wavevector $k=\frac{2n\pi}{N_s}$ with $n=1,2,.....,N_s$; 
filling-fraction  ${\rm f}\equiv \langle n_j\rangle =\frac{N_p}{N_s}$ with $N_p$ being
 the total number of HCBs in the system.
\begin{figure}
\includegraphics[width=0.6\linewidth,angle=-90]{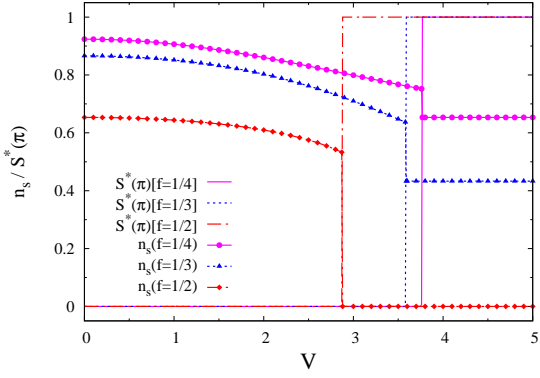}
\caption{(Color online) Plots of rescaled structure factor $S^*(\pi)$ and superfluid fraction
$n_s$ at various filling factors $\rm f$ obtained using modified Lanczos technique. The calculations were
 at   $\rm f = 1/2, 1/4$ with system size $N_s=16$ and
at ${\rm f}=1/3$ with  $N_s=12$. At a critical repulsion
there is a striking discontinuous transition in both $S^*(\pi)$ and $n_s$; while $S^*(\pi)$ jumps
from its minimum to maximum,  there is a significant drop in $n_s$. 
}
\label{ss_fig}
\end{figure}
{For e-rings with $N_s=2N$ sites, from the definition of the structure factor, for $k=\pi$ we have
\begin{align}
S(k)&=
\sum_{l=1}^{2N}(-1)^l W(l) .
\end{align}
Substituting the expression for the correlation function
$ W(l)$ and recognizing that $\sum_{l=1}^{2N}e^{i\pi l} =0 $,
we get
\begin{align}
S(k)=\frac{4}{2N}\sum_{j=1}^{2N}\Big \langle n_j\sum_{l=1}^{2N} n_{j+l} (-1)^l  \Big \rangle .
\end{align}
In the above equation, on taking $j+l=m$ we get 
\begin{align}
S(k)=\frac{2}{N}&\sum_{j=1}^{2N} \Big \langle n_j (-1)^j 
 \sum_{m=1}^{2N} n_{m} (-1)^m  \Big\rangle .
\end{align}
On defining the number operator which gives the total number of HCBs at even (odd) sites as
$\hat N_e=\sum\limits_{j_{\rm even}}n_j$  $(\hat N_o=\sum\limits_{j_{\rm odd}}n_j)$, we obtain  
\be
S(\pi)=\frac{2\langle(\hat N_e-\hat N_o)^2\rangle}{N} .
\ee
Due to the presence of only next-nearest-neighbor hopping, both $\hat N_e$ and $\hat N_o$ commute with the $t_2-V$ Hamiltonian;
 hence we obtain the following result
\be
S(\pi)=\frac{2( N_e- N_o)^2}{N} ,
\ee}
where $N_e$ ($N_o$ ) are the total number of HCBs at even (odd)
sites.
Thus the minimum value of $S(\pi) =0 $ corresponds to equal number of particles
in both the sub-lattices whereas the maximum is given by
\begin{eqnarray}
\left[S(\pi)\right]_{\rm max}=\frac{2N^2_p}{N}=\frac{4N^2_p}{N_s} ,
\label{sk_e}
\end{eqnarray}
indicating a single sublattice occupancy.
To study the QPT, we rescale the value of $S(\pi)$ as 
$S^*(\pi)= \frac{S(\pi)}
{\left[S(\pi)\right]_{\rm max}}$
with $S^*(\pi)$ representing the order parameter that varies
from 0 to 1 during the phase transition.

\begin{figure}
\includegraphics[width=0.5\linewidth,angle=-90]{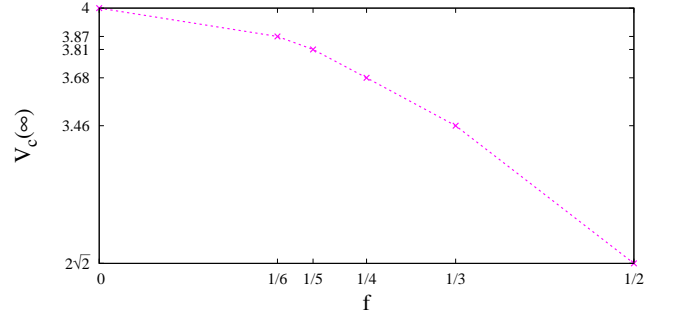}
\caption{(Color online) Plot of  $V_c(\infty)$ (critical repulsion for an infinite system)
obtained from Green's function analysis for half-filled (${\rm f} =1/2$) and two HCB systems (${\rm f} \rightarrow 0$)
and from finite size scaling at various other fillings $\rm f$.
}
\label{V_crit}
\end{figure}

Next, we will outline our procedure for calculating the superfluid density by
threading the chain with an infinitesimal magnetic flux $\theta$. 
The superfluid fraction is given by \cite{fisher,sdsy1}
\be
\!\!\!\!\!\!\!\!
n_s = \frac{N^2}{N_p t_{eff}}\left [ \frac{1}{2} \frac{\partial ^2 E(\theta)}{\partial \theta^2} \right]_{\theta =0}  ,
\label{ns}
\ee
where $E(\theta)$ is the total energy when threaded by flux $\theta$ and $t_{\rm eff} = \hbar^2/2m$ is the effective
hopping term which for our $t_2-V$ model is given by $t_{\rm eff}= 4t_2$.

The total energy for the case $V> V_c$, when threaded by a flux $\theta$,
 is expressed as
\be
E(\theta) = -2t_2\sd_{k} \cos[2(k+\theta/N_s)] ,
\ee
where $N_s=2N$.
Then, from the above definition of superfluid density $n_s$,
 for $V>V_c$, we have
\be
n_s=\frac{1}{N_p}\sum_k \cos (2k) .
\label{n_s}
\ee

Since we consider even values of $N_p$, the momenta occupied by the HCBs are $k=\frac{(2m+1)\pi}{2N}$ with $-\frac{N_p}{2}\le m \le \frac{N_p}{2}-1$.
Summing over these momenta, for the case of single sublattice occupancy (which occurs when $V > V_c$), we have from expression (\ref{n_s})
\be
n_s = \frac{1}{N_p} \frac{\sin \left ( \frac{ \pi N_p}{N} \right )}{\sin \left ( \frac{ \pi}{N} \right )} .
\label{ns1}
\ee
When both the sublattices are equally occupied (i.e., for $V=0$), in each sublattice of $N$ sites we have $\frac{N_p}{2}$ particles.
 Thus, the superfluid density in this case takes the form 
\be
n_s = \frac{2}{N_p} \frac{\sin \left ( \frac{ \pi N_p}{2N} \right )}{\sin \left ( \frac{ \pi}{N} \right )} .\label{ns2}
\ee

When e-rings
 were used, at all fillings ${\rm f}$, we found that the order parameter $S^*(\pi)$
jumps from 0 to 1 at a critical value of repulsion $V_c$ indicating that the system
transits from equally populated sub-lattices (i.e., Ising symmetry) case to a single sub-lattice
occupancy, i.e., a period-doubling CDW state [see Fig. \ref{ss_fig}]. Concomitantly, as can be seen from Fig. \ref{ss_fig},
there is a sudden drop in the superfluid fraction $n_s$
at the same critical repulsion. At half-filling, where the superfluid fraction vanishes
above a critical repulsion because a single sub-lattice is completely filled, the transition shows that 
superfluidity and CDW state are mutually exclusive.
On the other hand, at all non-half-fillings, we see that the system undergoes a QPT from a superfluid
to a supersolid (i.e., a homogeneously coexisting superfluid and CDW) state.
In Fig. \ref{ss_fig}, it is of interest to note that the values of $n_s$ at $V=0$ and $V > V_c$
are exactly those predicted by Eq. (\ref{ns1}) and Eq. (\ref{ns2}), respectively.

Next, for e-rings at various fillings,
we relate $V_c(2N)$ (critical repulsion at $N_s=2N$) to  $V_c(\infty)$ using finite size scaling analysis 
(see Appendix A for details)
and obtain Fig. \ref{V_crit}.
 At half-filling, from finite size  scaling analysis we obtain that $V_c(\infty) \approx 2.83$;
in the next section, we show (using Green's function analysis)  the exact result $V_c(\infty) = 2\sqrt{2}$.
For systems with 2 HCB and $N_s=4,6,8,10,12,14,16,18,$ and $20$, we find numerically  that $V_c \approx 4.00$; in
the next section, we obtain the exact result (using Green's functions)
that $V_c(2N) =4$ for any system size $N_s=2N\ge 4$.


\section{Exact instability condition at half-filling in \lowercase{e-rings}}\label{sec3}
We study the following $t_2-V$ Hamiltonian
in rings with even number of sites ($2N$) by considering two sublattices C and D
and using periodic boundary conditions:
\begin{eqnarray}
H \equiv &&  
 - t_2 \sum_{i=1} ^{N} (c^{\dagger}_{i} c_{i+1} + {\rm H.c.} ) 
- t_2 \sum_{i=1}^{N} (d^{\dagger}_{i}d_{i+1}  + {\rm H.c.} ) 
\nonumber \\
&&
+V \sum_{i=1}^{N} d^{\dagger}_i d_{i} (c^{\dagger}_i c_i + c^{\dagger}_{i-1} c_{i-1}) ,
\label{eq:tV}
 \end{eqnarray}
where $c$ and $d$ denote destruction operators of HCBs in
sublattices $C$ and $D$, respectively.

To understand the 
discontinuous phase transition at half-filling,
(i.e., the transition from  equal occupation of both sublattices  to occupation of only one sublattice
at a critical $V_c$)
we begin by recognizing that when the system is on the verge of
completing the phase transition, the system will pass through the state
where there is one HCB in one sublattice and one hole in the other sublattice.
Hence, we now consider instability for the case
of one particle in sublattice $C$ and one hole [with destruction operator denoted by $h$ ($\equiv d^{\dagger}$)]
in sublattice $D$; we then rearrange the above equation as
\begin{eqnarray}
H \equiv &&  
 - t_2 \sum_{i=1}^{N} (c^{\dagger}_{i} c_{i+1} + {\rm H.c.} ) 
+ t_2 \sum_{i=1}^{N} (h^{\dagger}_{i}{h}_{i+1}  + {\rm H.c.} ) 
\nonumber \\
&&
-V \sum_{i=1}^{N} (h^{\dagger}_i h_{i}-1) (c^{\dagger}_i c_i + c^{\dagger}_{i-1} c_{i-1}) .
\label{eq:thV}
 \end{eqnarray}
We define the particle-hole Green's function as follows \cite{berciu2}:
\begin{eqnarray}
 g^h_n \equiv {_h \!}\langle k,0|G(\omega)|k,n \rangle_{h} ,
\end{eqnarray}
where 
$G(\omega) \equiv 1/ (\omega +i\eta -H) $
 and $|k,n\rangle_h$ is the particle-hole state 
\begin{eqnarray}
|k,n \rangle_h = \frac{1}{\sqrt{N}} \sum_{j=1}^{N} e^{ik \left (j+\frac{n}{2}\right )} c^{\dagger}_j h^{\dagger}_{j+n} |0 \rangle ,
\end{eqnarray}
with $k$ being the total momentum of the particle-hole system and $n$ the separation between the particle and the hole.
Using the condition
\begin{eqnarray}
 \delta_{0,n} \equiv {_h\!}\langle k,0|G(\omega)(\omega +i\eta -H)|k,n \rangle_h ,
\label{delta}
\end{eqnarray}
for $n=0$, we obtain
 \begin{eqnarray}
 (\omega +i\eta)g^h_{0} &=&1+{_h\!}\langle k,0|G(\omega)(H)|k,0 \rangle_h
\nonumber \\
&=& 1 + Vg^h_0-i 2t_2\sin({k}/2)g^h_1 
\nonumber \\
&&+i2t_2 \sin({k}/{2})g^h_{-1} .
\label{g0}
\end{eqnarray}
From Eq. (\ref{delta}), for $n=1$, we get 
\begin{eqnarray}
 (\omega +i\eta)g^h_{1} &=&{_h\!}\langle k,0|G(\omega)(H)|k,1 \rangle_h
\nonumber \\
&=&  Vg^h_1-i 2t_2\sin({k}/2)g^h_2 
\nonumber \\
&&+i2t_2 \sin(k/2)g^h_{0} .
\label{g1}
\end{eqnarray}
Similarly, for $n \neq 0,1$, we derive
\begin{eqnarray}
 (\omega +i\eta)g^h_{n} &=&{_h\!}\langle k,0|G(\omega)(H)|k,n \rangle_h
\nonumber \\
&=&  2Vg^h_n-i 2t_2\sin({k}/2)g^h_{n+1} 
\nonumber \\
&&+i2t_2 \sin(k/2)g^h_{n-1} .
\label{g_n}
\end{eqnarray}

As $V$ is increased to the critical repulsion $V_c$, the  HCB in sublattice $C$  
vacates its sublattice and enters the sublattice $D$ containing the hole; the energy of
the system then becomes 0.
Next, we let
\begin{eqnarray}
\frac{\omega+i\eta-2V}{F_k} \equiv \frac{1}{z}-z ,
\label{eqnz}
\end{eqnarray}
where $F_k \equiv i2t_2\sin(k/2)$.
Then, Eq. (\ref{g_n}) takes the simple form
\begin{eqnarray}
 \left ( \frac{1}{z}-z \right ) g^h_n =g^h_{n-1}-g^h_{n+1} ,
\label{grecur}
\end{eqnarray}
whose solution is of the form
\begin{eqnarray}
 g^h_n = \alpha^{\pm}_1 z^n + \frac{\beta^{\pm}_1}{(-z)^n} ,
\label{gzn}
\end{eqnarray}
where $\alpha^+_1$ ($\alpha^-_1$) and $\beta^+_1$ ($\beta^-_1$) correspond to $n >1$ ($n<0$). 
The transition occurs at the critical value of $V$ that makes the overall energy 0.
Let $V =2t_2 \gamma \sin(k/2) $. The overall energy is less than $-4t_2 +2V$ (for $V > 0$).
 It is important to note that, for the groundstate, $k=\pi$ for any $V\le V_c$. This can be seen by first noting
that when $V=0$, total
momentum in the minimum energy state is $\pi$; next, turning on $V$ does not change
the total momentum.
Then for $k=\pi$, to get overall energy to be 0, we need the inequality $\gamma > 1$.
The instability condition corresponds to the case $\omega =0$
because then the Green's function $g^h_0$ diverges when the energy is zero.
It then follows from Eq. (\ref{eqnz}) that 
\begin{eqnarray}
 i 2 \gamma = \frac{1}{z}-z ,
\end{eqnarray}
which implies that
\begin{eqnarray}
 z=i(-\gamma +\sqrt{\gamma^2 -1}) ,
\end{eqnarray}
and hence $|z| < 1$ for $\gamma > 1$.
Let us first consider the case $n > 1$.
From the above Eq. (\ref{gzn}), it is clear that, for $n \rightarrow \infty$,
$g^h_n$ is finite only for $\beta^+_1 =0$. Thus for $n>1$,
\begin{eqnarray}
 g^h_{n+1}= z g^h_{n} ,
\label{gn+1}
\end{eqnarray}
 which 
implies that $g^h_3 = z g^h_2$; then, from Eq. (\ref{grecur}) it follows that
\begin{eqnarray}
g^h_2 = z g^h_1 . 
\label{g2}
\end{eqnarray}
Next, for the case $n < 0$,
we see that $g^h_n$ is finite, for $n \rightarrow -\infty$, only when $\alpha^-_1 =0$.
Thus for $n < 0$,
\begin{eqnarray}
 g^h_{n-1}= -z g^h_n ,
\label{gn-1}
\end{eqnarray}
 which implies that $g^h_{-2} = -z g^h_{-1}$;
then from Eq. (\ref{grecur}) we obtain 
\begin{eqnarray}
 g^h_{-1} = -z g^h_0 .
\label{g-1}
\end{eqnarray}
From Eqs. (\ref{g0}), (\ref{g1}), (\ref{g2}), and (\ref{g-1}), we obtain
\begin{eqnarray}
 g^h_0 = \frac{1}{(\omega +i\eta - V + z F_k ) + \frac{F_k^2}{(\omega +i\eta - V + z F_k )}} ,
\label{g0div}
\end{eqnarray}
and
\begin{eqnarray}
 g^h_1 = \frac{g^h_0 F_k}{(\omega +i\eta - V + z F_k ) } ,
\label{g1div}
\end{eqnarray}
where $\omega =0$, $V=2t_2 \gamma$, $z=i(-\gamma +\sqrt{\gamma^2 -1})$, and $F_k =i2t_2$.
It then follows that $g^h_0$ diverges when $(V-zF_k)^2 +F_k^2 =0$, i.e., when $\gamma = \sqrt{2}$.
Thus the instability condition is $V_c = 2 \sqrt{2}t_2$. (We now see that the total energy at transition
is indeed less than $-4t_2 + 2V$).
It is important to note [as can be seen from Eq. (\ref{g1div})] that, when $g^h_0$ diverges,  $g^h_1$ also diverges;
consequently, from Eqs. (\ref{gn+1}), (\ref{gn-1}), and (\ref{g-1})  we see that all $g^h_n$ diverge (i.e., even when $n >1$ and $n < 0$).


\section{Exact instability condition for two HCB\lowercase{s} in \lowercase{e-rings}}\label{sec4}
 We now study the non-trivial case of the two-HCB instability for the $t_2-V$ Hamiltonian [described by Eq. (\ref{eq:tV})]
in e-rings.
 We consider one HCB in sublattice C and one HCB in
sublattice D; the corresponding two-particle Green's function is
defined as
$ g_n \equiv \langle k,0|G(\omega)|k,n \rangle $
with
$G(\omega)$ being defined as before 
 and the two-particle  state $|k,n\rangle$ being expressed as
\begin{eqnarray}
|k,n \rangle = \frac{1}{\sqrt{N}} \sum_{j=1}^{N} e^{ik \left (j+\frac{n}{2}\right )} c^{\dagger}_j d^{\dagger}_{j+n} |0 \rangle ,
\end{eqnarray}
with  $k$ representing the total momentum of the two-particle system.
Then, the following equations hold for the Green's functions $g_n$:
\begin{eqnarray}
 (\omega +i\eta -V)g_0 &=& 1 -2t_2\cos({k}/2)g_1 
\nonumber \\
&&-2t_2 \cos(k/2)g_{-1} ,
\label{g02hcb}
\end{eqnarray}
\begin{eqnarray}
 (\omega +i\eta -V)g_1 &=& -2t_2\cos( k/2)g_2 
\nonumber \\
&& -2t_2 \cos(k/2)g_0  ,
\label{g12hcb}
\end{eqnarray}
and for $n\neq 0,1$
\begin{eqnarray}
(\omega +i\eta )g_n &=& -2t_2\cos(k/2)g_{n+1} 
\nonumber \\
&&-2t_2 \cos(k/2)g_{n-1} .
\label{gn}
\end{eqnarray}

As $V$ increases to the critical
$V_c$, the energy given by Eq. (\ref{eq:tV}) 
becomes $-4t_2\cos(\pi/N)$ (i.e., the minimum energy of the two HCBs in the same sublattice);
this would correspond to the instability 
where one HCB 
quits its sublattice and goes into the sublattice of the other particle. 
Here, we make the key observation that $k=0$ for the groundstate at any V. To understand this,
we first note for $V=0$, the total
momentum is zero in the minimum energy state; next, we recognize that turning on $V$ does not change
the total momentum.
Now, to obtain the instability, we take
\begin{eqnarray}
\frac{\omega}{(-2t_2)} &=& \frac{4t_2\cos(\pi/N)}{2t_2} = 2\cos(\pi/N)
\nonumber \\
&=& e^{i\pi/N}+e^{-i\pi/N} 
\nonumber \\
&=& z+ \frac{1}{z} .
\end{eqnarray}
We set $z=e^{i\pi/N}$
and  also take $V/(2t_2) =2 \gamma$.Then, Eqs. (\ref{g02hcb}), (\ref{g12hcb}), and (\ref{gn})
become
\begin{eqnarray}
 [(z+1/z) +2 \gamma]g_0 = -1/(2t_2) +g_1 + g_{-1} ,
\label{g0_2}
\end{eqnarray}
\begin{eqnarray}
 [(z+1/z) +2 \gamma]g_1 = g_2 + g_{0} ,
\label{g1_2}
\end{eqnarray}
and for $n\neq 0,1$
\begin{eqnarray}
 [(z+1/z) ]g_n = g_{n+1} + g_{n-1} .
\label{gn_2}
\end{eqnarray}
Without loss of generality, we assume
\begin{eqnarray}
 g_1 = \alpha_2 z + \beta_2/z ,
\label{g1_z}
\end{eqnarray}
and
\begin{eqnarray}
 g_2 = \alpha_2 z^2 + \beta_2/z^2 .
\label{g2_z}
\end{eqnarray}
Then using Eqs. (\ref{gn_2}), (\ref{g1_z}), and (\ref{g2_z}), we obtain for $n=3,4,...,N$ the expression
\begin{eqnarray}
 g_n = \alpha_2 z^n + \beta_2/z^n .
\label{gn_z}
\end{eqnarray}
It is important to recognize that $|z| =1$; hence, the
Green's functions do not decay with HCB separation $n$. Next, we  note that
at $k=0$, $|k,-n\rangle = |k,N-n\rangle$; consequently, we
see that $g_{N} =g_0$ and $g_{N-1} = g_{-1}$.
Then, using Eq. (\ref{gn_z})
and the relation $z=e^{i\pi/N}$, we get
\begin{eqnarray}
g_{-1} = g_{N-1} = \alpha_2 z^{N-1} + \beta_2/z^{N-1} 
\nonumber \\
=-(\alpha_2/z +\beta_2 z),
\label{g2_N-1}
\end{eqnarray}
and 
\begin{eqnarray}
 g_0 = g_{N} = \alpha_2 z^{N} + \beta_2/z^{N} 
\nonumber \\
=-(\alpha_2 +\beta_2 ).
\label{g2_N}
\end{eqnarray}

We are now ready to solve for the Green's functions $g_n$ using Eqs. (\ref{g0_2}), (\ref{g1_2}), (\ref{g1_z}),
(\ref{g2_z}), (\ref{g2_N-1}), and (\ref{g2_N}). We get the following equations:
\begin{eqnarray}
 \alpha_2[z+\gamma ] + \beta_2 [ 1/z +\gamma] = 1/(4t_2) ,
\label{alpha1}
\end{eqnarray}
and
\begin{eqnarray}
 \alpha_2[1+\gamma z] + \beta_2 [ 1 + \gamma /z] =0 .
\label{alpha2}
\end{eqnarray}
It then follows from the above two equations that
\begin{eqnarray}
 \alpha_2= \frac{z+\gamma}{4t_2(z^2 + \gamma^2 -1 - \gamma^2 z^2)} ,
\label{alpha}
\end{eqnarray}
which diverges when $\gamma = \pm1$. We get $\gamma =1$ for repulsive $V$.
Furthermore,
\begin{eqnarray}
 \beta_2= -\alpha_2 \frac{1+\gamma z}{1+\gamma/z} ,
\label{beta_alpha}
\end{eqnarray}
which for $\gamma =1$ yields $\beta_2 = -\alpha_2 z$.
Thus we see that $g_0 =-(\alpha_2+\beta_2)=-\alpha_2(1-z)$ diverges for $\gamma=1$ or $V_c=4t_2$.
We also find that for $0\neq n \le N $ 
\begin{eqnarray}
\!\!\!\!\!\!\!\! g_n = \alpha_2 z^n + \beta_2/z^n = \alpha_2 \left ( e^{i\pi n/N} - e^{-i \pi (n-1)/N} \right )  ,
\end{eqnarray}
also diverges since $2n-1 \neq 2N$.
{\em The instability condition $V_c = 4t_2$ is independent of N and hence is valid 
for large $N$ as well!}
Another interesting observation based on $\beta_2 = -\alpha_2 z$ is that 
\begin{eqnarray}
g_{-k}=g_{N-k} &=& - \alpha_2/ z^k - \beta_2 z^k
\nonumber \\
& =& \beta_2 /z^{k+1} + \alpha_2 z^{k+1} 
\nonumber \\
&=& g_{k+1} .  
\end{eqnarray}


\begin{figure}[b]
\includegraphics[width=0.6\linewidth,angle=-90]{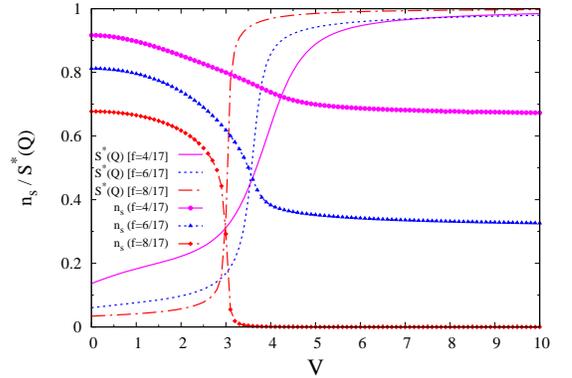}
\caption{(Color online) Plots of rescaled structure factor $S^*(Q) \equiv {S(Q)}/
{\left[S(Q)\right]_{\rm max}} $ (with $Q=\pi\pm\frac{\pi}{2N+1}$) and superfluid fraction
$n_s$ at various fillings $\rm f$ obtained using modified Lanczos method. At a critical repulsion
there is a sharp rise in
$S^*(Q)$ with a concomitant 
significant drop in $n_s$. 
}
\label{ss_fig2}
\end{figure}

\begin{figure}[h]
\includegraphics[width=0.7\linewidth,angle=-90]{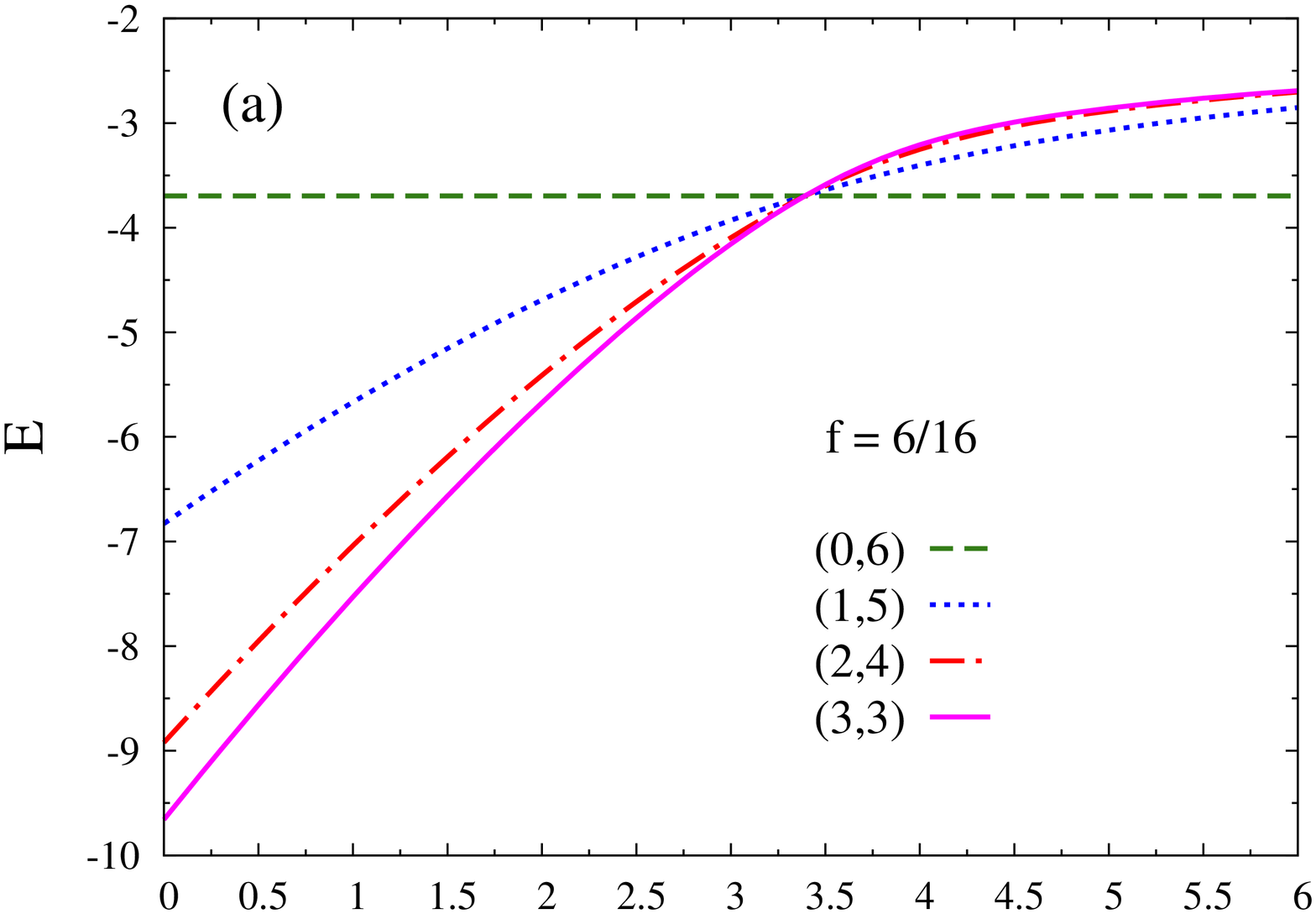}
\includegraphics[width=0.7\linewidth,angle=-90]{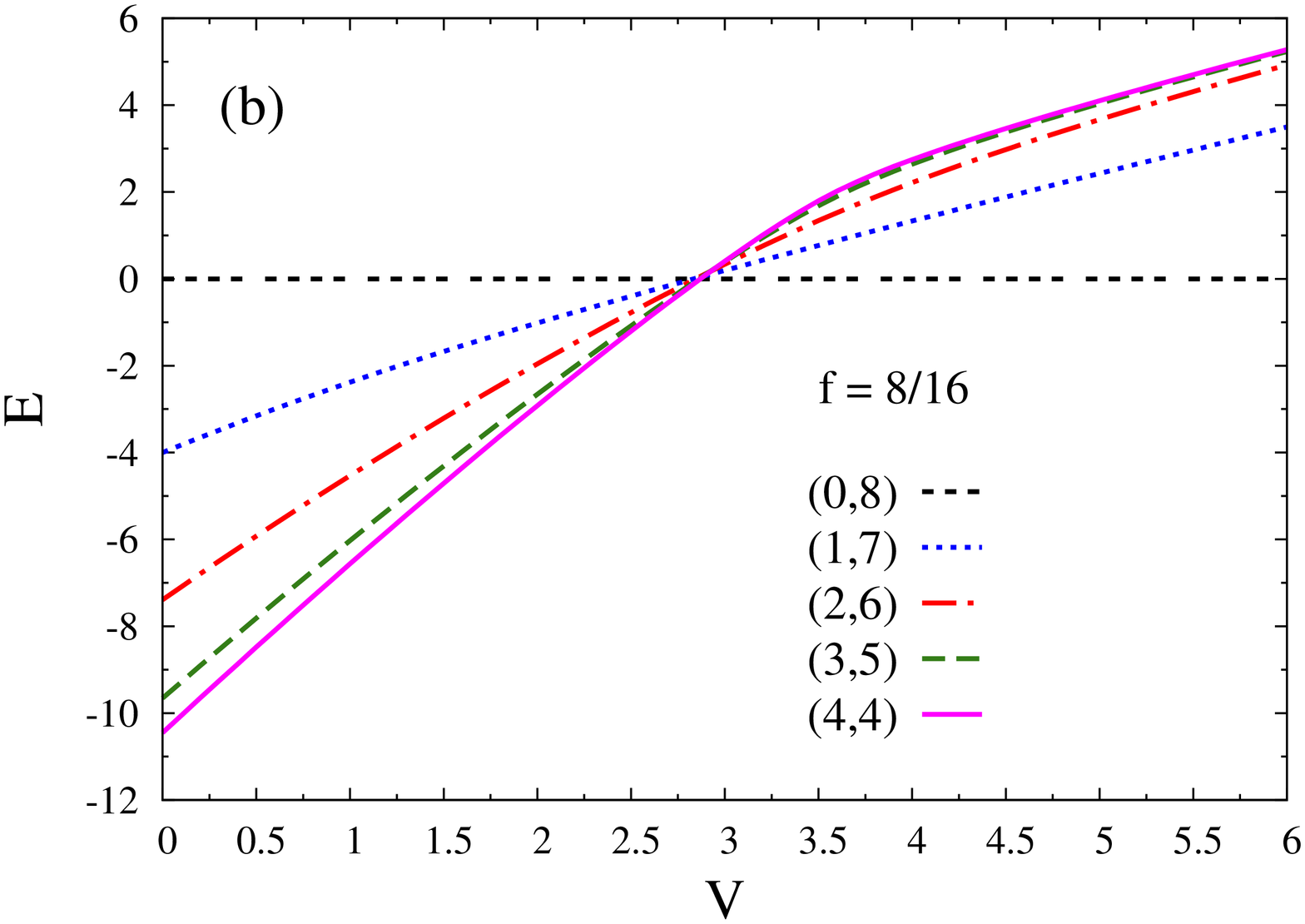}
\caption{(Color online) Plots of energy E versus repulsion V for (a) the 
lowest energy states with (m,6-m) particles in e-ring with $N_s=16$ and $N_p=6$ (non-half filled case); 
and (b) the lowest energy states with (m,8-m) particles in e-ring with $N_s=16$ and $N_p=8$ (half-filled case).
}
\label{e-rings_fig}
\end{figure}


\section{Numerical study of the $t_2-V$ model for \lowercase{o-rings}}\label{sec5}
We calculate the structure factor at large repulsion so that the CDW would be better defined (with larger
values for the structure factor)
even for finite number of sites.
Since the allowed momenta for HCBs in o-rings (with $2N+1$ sites) are $k=\frac{2n\pi}{2N+1}$ with n=1,2,.....,2N+1; 
the structure factor $S(k)$ will have peaks at $k=Q\equiv \pi\pm\frac{\pi}{2N+1}$; in fact $\pi$ is not an allowed value for the momentum k!.
For $k=Q=\pi+\frac{\pi}{2N+1}$ we have
\begin{align}
S(Q)&=\sum_{l=1}^{2N+1}e^{i\left(\pi+\frac{\pi}{2N+1}\right)l}W(l)\nonumber\\
&=\sum_{l=1}^{2N+1}(-1)^le^{i\left(\frac{\pi}{2N+1}\right)l}W(l) .
\end{align}
Substituting the expression for $W(l)$
and ignoring the term $\langle n_j\rangle\langle n_{j+l}\rangle=\langle n_j\rangle^2$ in
$W(l)$ (because $\sum_{l=1}^{2N+1}e^{i\left(\pi+\frac{\pi}{2N+1}\right)l} =0 $)
we get
\begin{align}
S(Q)=\frac{4}{2N+1}\sum_{j=1}^{2N+1}\Big \langle n_j\sum_{l=1}^{2N+1} n_{j+l} (-1)^l e^{i\left(\frac{\pi}{2N+1}\right)l} \Big \rangle .
\end{align}
In the above equation, setting $j+l=m$ yields 
\begin{align}
S(Q)=\frac{4}{2N+1}&\sum_{j=1}^{2N+1} \Big \langle n_j (-1)^j e^{-i\left(\frac{\pi }{2N+1}\right)j}\nonumber\\
&~~ \times \sum_{m=1}^{2N+1} n_{m} (-1)^m e^{i\left(\frac{\pi}{2N+1}\right)m} \Big\rangle .
\end{align}
At large repulsion all the particles in o-rings will be confined to $N$ alternate sites similar to the case
of e-rings
(with $2N$ sites). First, we consider the filling ${\rm f} = N/(2N+1)$ (i.e., $N_p=N$). 
 Consequently, the maximum value of the structure factor is given by
\begin{align}
[S(Q)]_{\rm max}=& \frac{4}{2N+1}\Big[\frac{N_p}{N}(-1)e^{-i\frac{\pi}{2N+1}}+\frac{N_p}{N}(-1)^3 e^{-i\frac{3\pi}{2N+1}}\nonumber\\
&~~~~~~+\cdots +\frac{N_p}{N}(-1)^{2N-1}e^{-i\frac{(2N-1)\pi}{2N+1}}\Big]\Big[{\rm H.c.}\Big]\nonumber\\
=&\frac{2}{2N+1} \left(\frac{N_p}{N}\right)^2 \left(\frac{1}{1-\cos\left(\frac{\pi}{2N+1}\right)}\right) ,
\label{sk_o}
\end{align}
with $N_p=N$ for ${\rm f} = N/(2N+1)$. For the $N_p=N$ case, the above expression for $[S(Q)]_{\rm max}$ is exact at all values of $N$ (see Table \ref{table_sf1}).

In the thermodynamic limit ($2N+1\rightarrow\infty$), the above expression for $[S(Q)]_{\rm max}$ diverges as
\begin{align}
[S(Q)]_{\rm max}=\frac{4}{\pi^2} \left(\frac{4 N_p^2}{2N+1}\right) = \frac{4}{\pi^2} \left(\frac{4 N_p^2}{N_s}\right) .
\label{sk_inf}
\end{align}

\begin{table}[b]
\begin{center}
{
\begin{tabular}{|c|c|c|c|c|}
\hline
\multicolumn {1}{|c|}{$N_p/N_s$} & {Analytical} &
\multicolumn{3}{|c|}{Numerical value of $[S(Q)]_{\rm max}$} \\ \cline{3-5} 
{} & {value of $[S(Q)]_{\rm max}$} & {V=50} & {V=100} & {V=500} \\
\hline
$\frac{4}{9}$ & $3.6848$ &  $3.6836$ & $3.6845$ & $3.6848$  \\
\hline
$\frac{6}{13}$ & $5.2944$ &  $5.2935$ & $5.2942$ & $5.2944$  \\
\hline
$\frac{8}{17}$ & $6.9095$  & $6.90878$ & $6.9093$ & $6.9095$  \\
\hline
$\frac{10}{21}$ & $8.5269$ & $8.5263$ & $8.5267$ & $8.5269$ \\
\hline
\end{tabular}}
\caption{{At filling fraction ${\rm f}=\frac{N}{(2N+1)}$,
 the numerical value of  $[S(Q)]_{\rm max}$ calculated at large $V$ agrees quite well with
the analytic value obtained from Eq. (\ref{sk_o}).
}}\label{table_sf1}
\end{center}
\end{table} 


\begin{figure}
\includegraphics[width=0.7\linewidth,angle=-90]{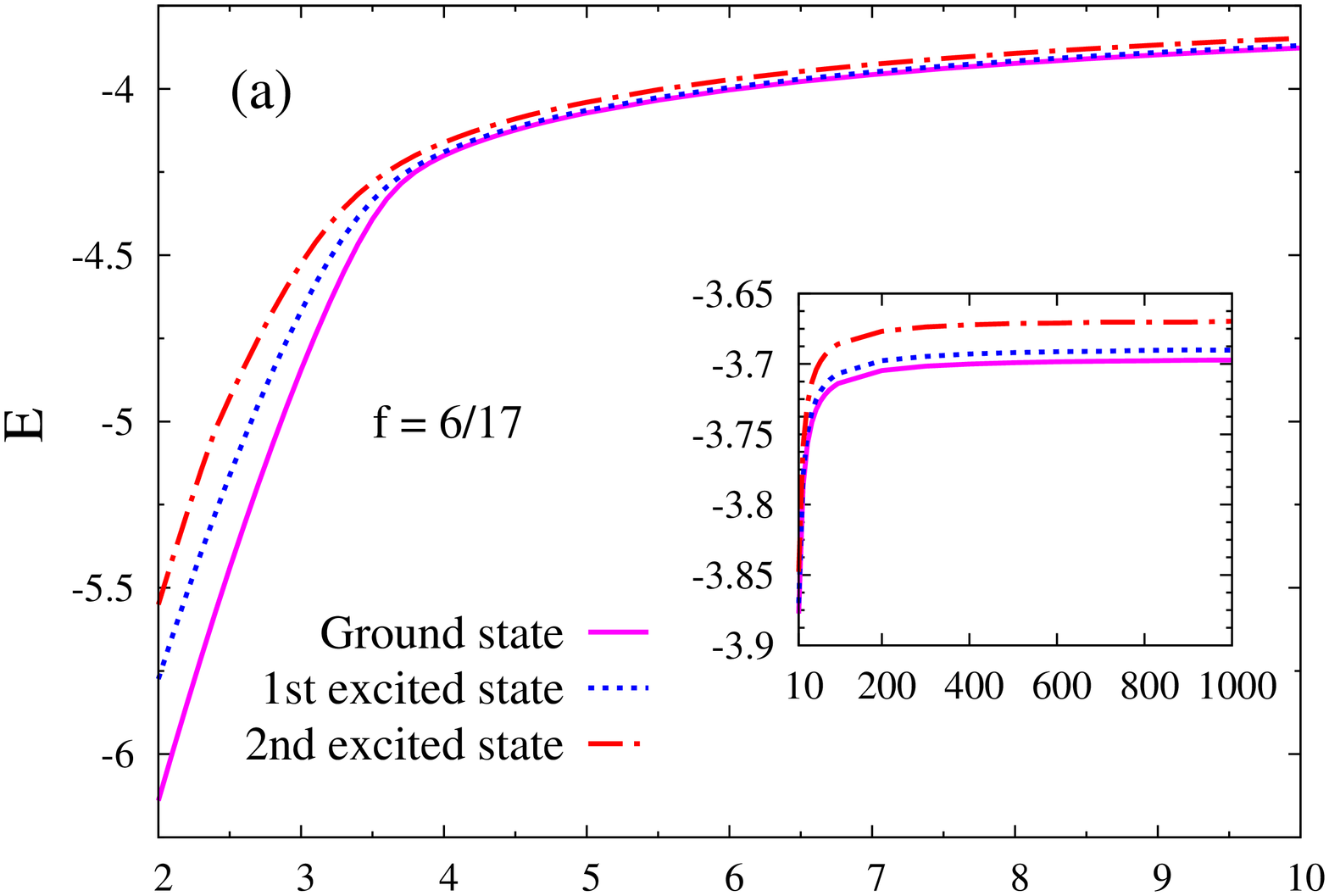}
\includegraphics[width=0.7\linewidth,angle=-90]{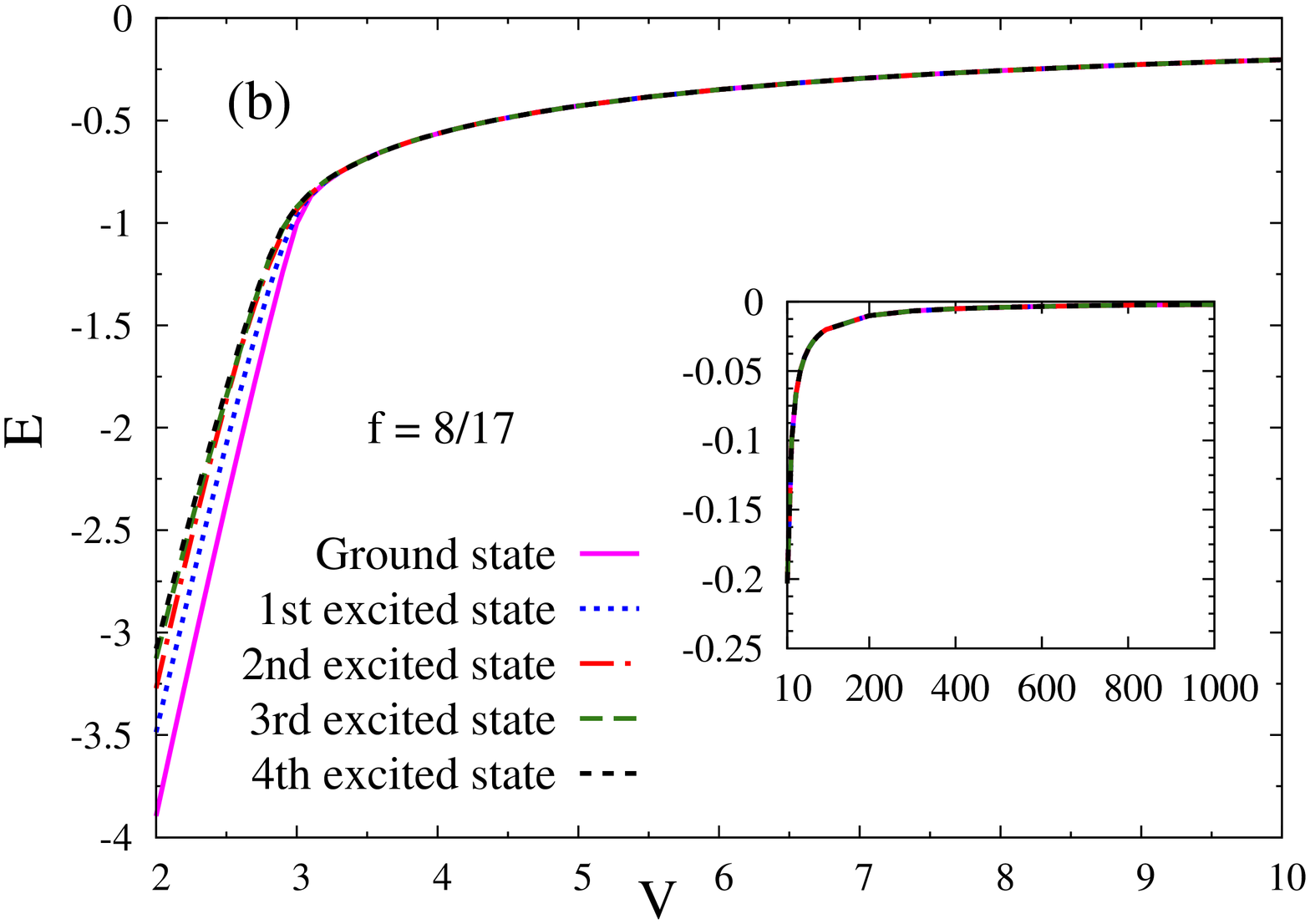}
\caption{(Color online) Plots of energy E versus repulsion V for (a) the three lowest energy states in 
o-ring with $N_s=17$ and $N_p=6$; and 
(b) the five lowest energy states in o-ring with $N_s=17$ and $N_p=8$.
}
\label{o-rings_fig}
\end{figure}


\begin{figure}[t]
\includegraphics[width=0.7\linewidth,angle=-90]{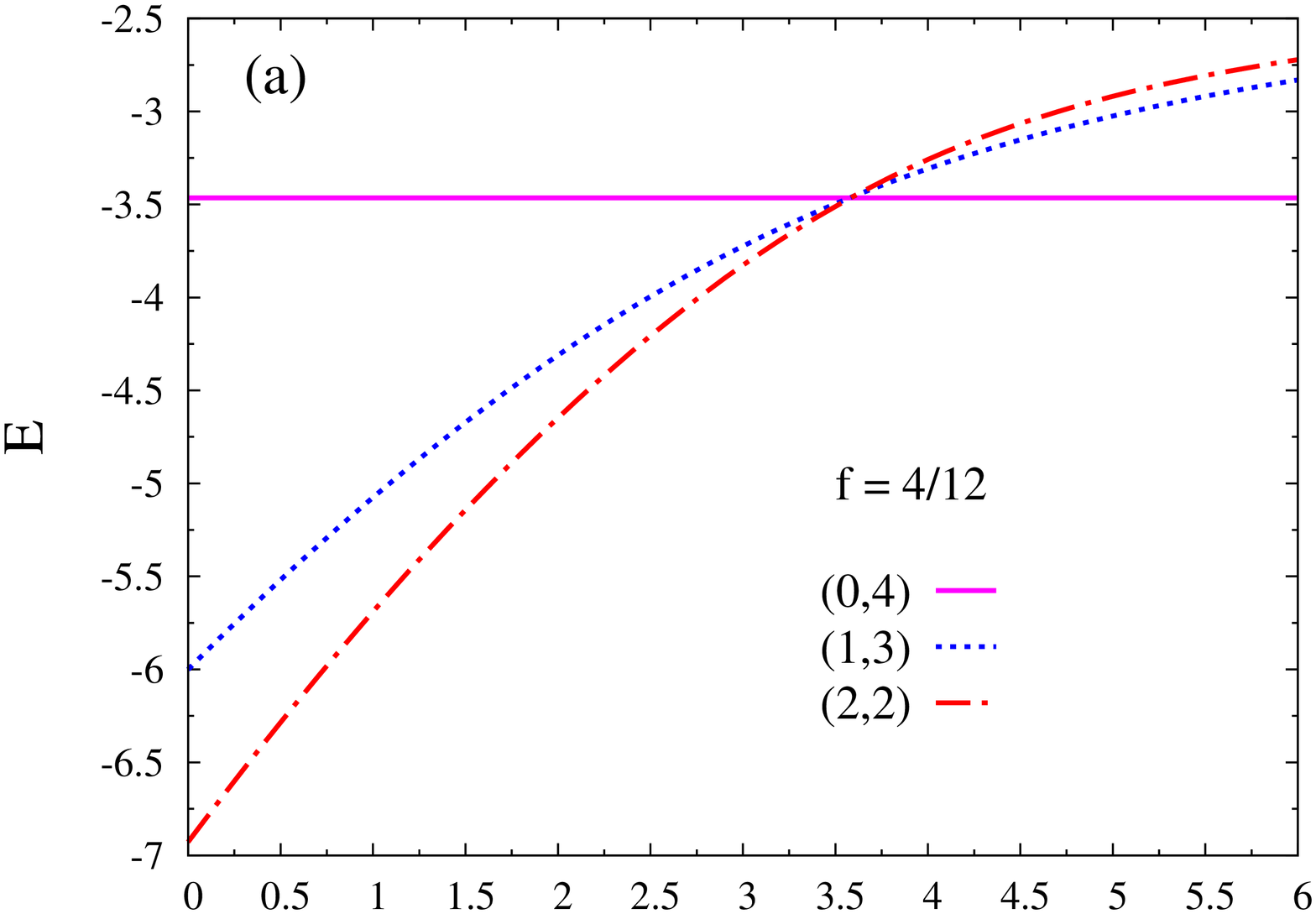}
\includegraphics[width=0.7\linewidth,angle=-90]{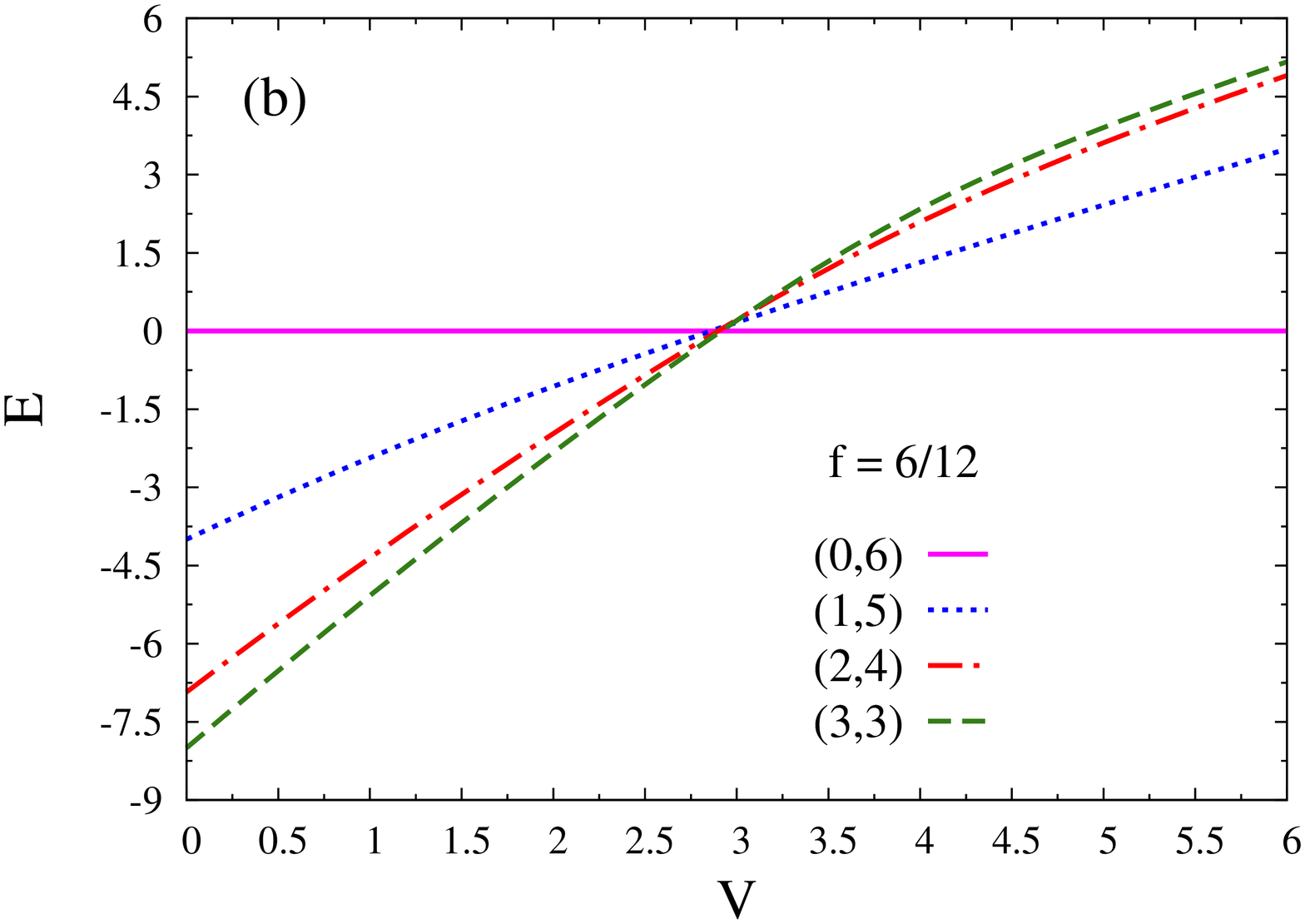}
\caption{(Color online) Plots of energy E versus 
repulsion V for (a) the lowest energy states with (m,4-m) particles in e-ring 
with $N_s=12$ and $N_p=4$ (non-half filled case); 
and (b) the lowest energy states with (m,6-m) particles in e-ring with $N_s=12$ and $N_p=6$ (half-filled case)
}
\label{e-rings_fig2}
\end{figure}

\begin{figure}[h]
\includegraphics[width=0.7\linewidth,angle=-90]{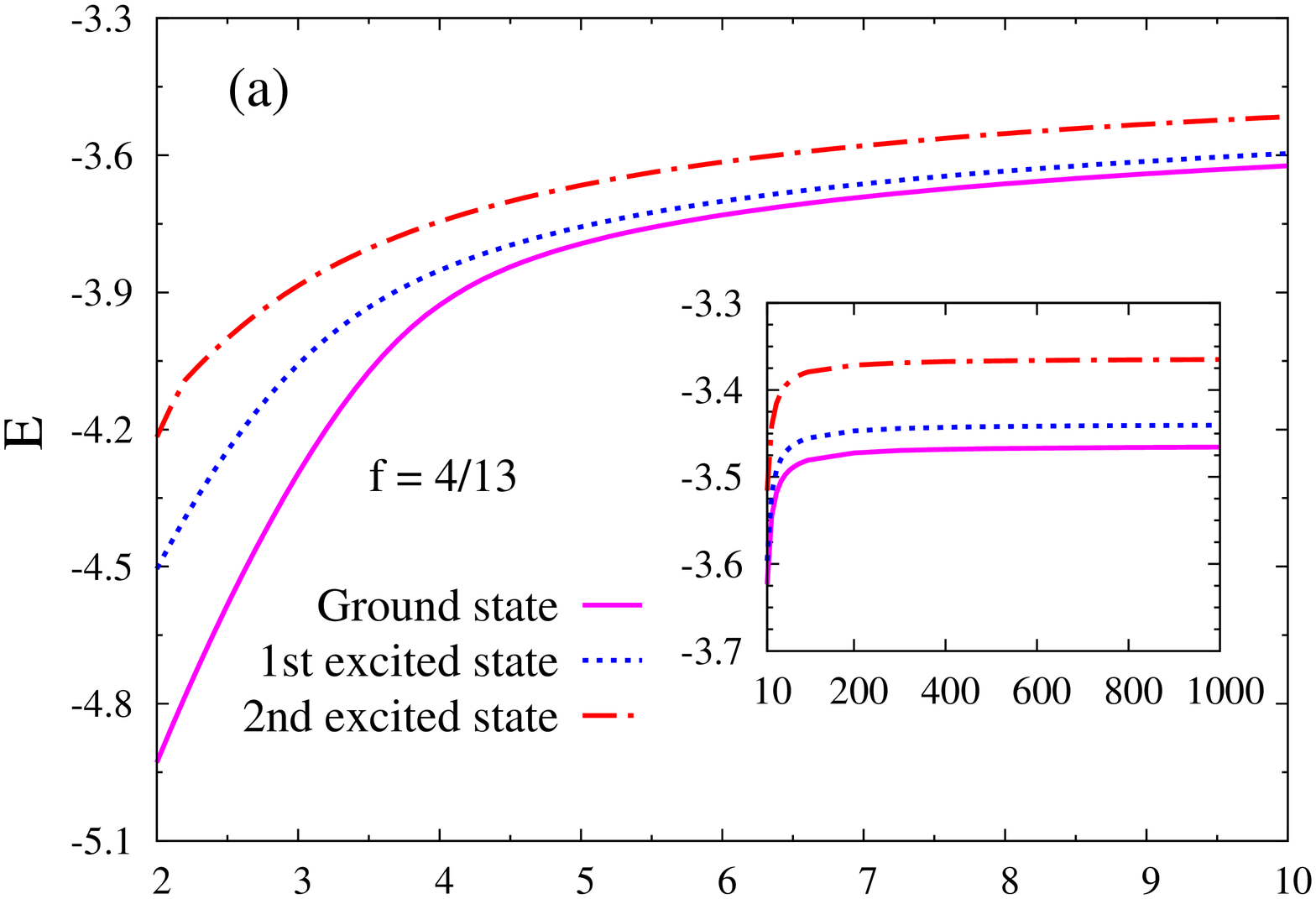}
\includegraphics[width=0.7\linewidth,angle=-90]{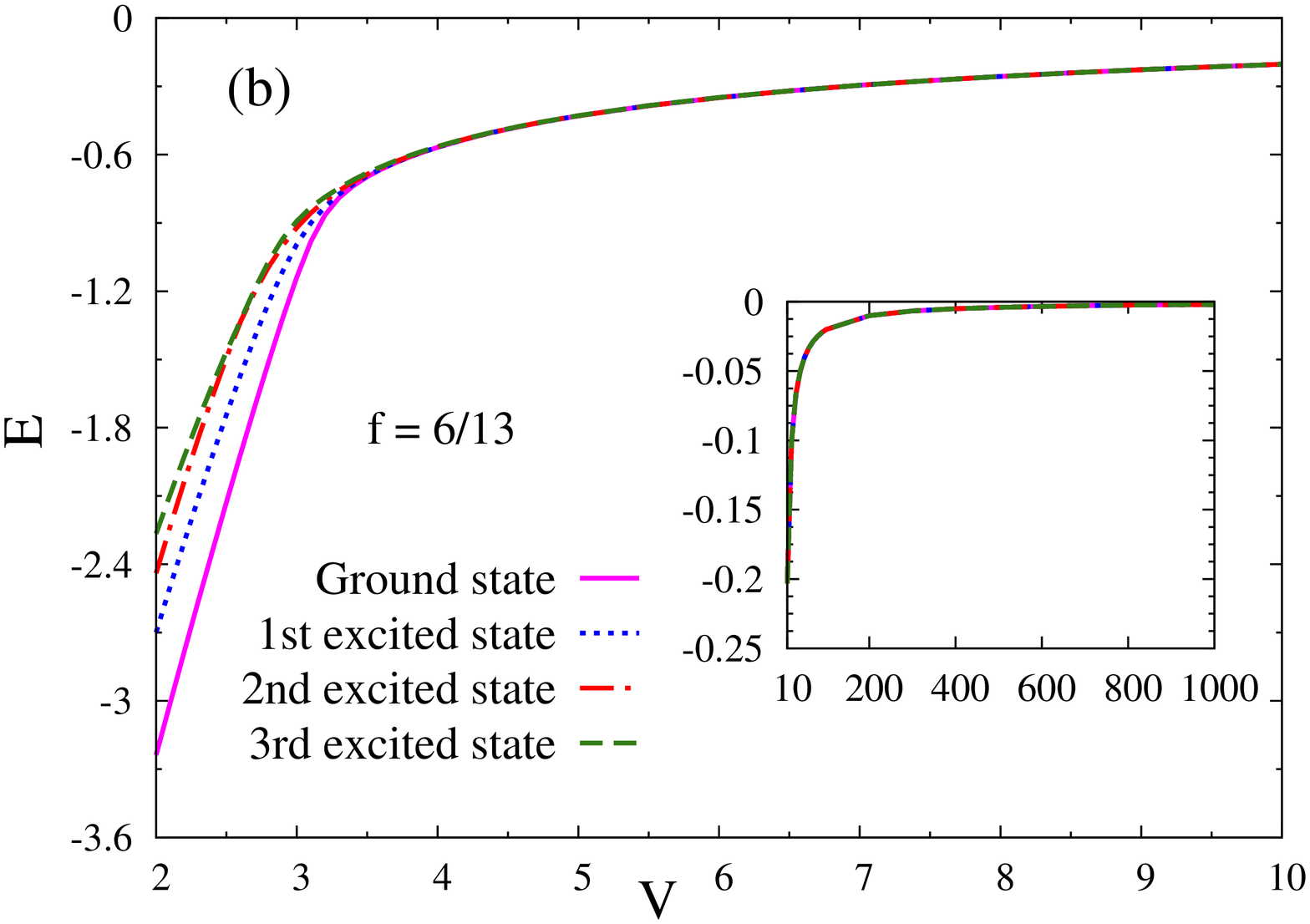}
\caption{(Color online) Plots of energy E versus repulsion V for (a) the three lowest 
energy states in o-ring with $N_s=13$ and $N_p=4$; 
and (b) the four lowest energy states in o-ring with $N_s=13$ and $N_p=6$.
}
\label{o-rings_fig2}
\end{figure}

We rescale  $S(Q)$ as 
\begin{align}
S^*(Q)=\frac{S(Q)}{[S(Q)]_{\rm max}}.
\end{align}
For $N_p < N$, all the $N$ sites considered for particle occupation would be connected through hopping so that the total 
energy is minimized, i.e., the potential energy is zero and the kinetic energy is minimized. Next, we assume that
the $N_p$ particles are uniformly distributed among these $N$ alternate sites; such an assumption
is valid for large values of $N$ as the end effects (i.e., at the boundary of the $N$ alternate sites)
is negligible.
Then, for $N_p < N$, the expression (\ref{sk_o}) is approximate at finite $N$ (see Table \ref{table_sf2}) and exact for $N \rightarrow \infty$.
Furthermore, the expression for the peak value of the structure factor for e-rings [given by Eq. (\ref{sk_e})]
differs from that for o-rings [given by Eq. (\ref{sk_inf})] due to the phase factor $e^{-i\left(\frac{\pi}{2N+1}\right)m}$
in Eq. (\ref{sk_o});  interestingly the difference is not negligible in the thermodynamic limit (i.e., $2N+1 \rightarrow \infty$).

Next, we will outline our procedure for calculating the
superfluid density for o-rings at $V = 0$.
From the definition of superfluid density $n_s$ it is obvious that, to calculate $n_s$, all we need is to calculate the total energy 
$E(\theta)$ which is independent of whether the particles are 
fermions or hard-core-bosons. 
So, if we recast our system in terms of fermions, the boundary condition turns out to be
\be
e^{ik(2N+1)2}=-1=e^{i(2m+1)\pi},
\ee
where $m$ is an integer. The above equation
implies that
\begin{eqnarray*}
k=\frac{(2m+1)\pi}{2(2N+1)} \quad{\rm with}\quad -\frac{N_p}{2}\leq m \leq \frac{N_p}{2}-1 .
\end{eqnarray*}

So, for $V=0$, we obtain the superfluid density for o-rings
\begin{align}
n_s&=\frac{1}{N_p}\sum\limits_{k=-\frac{\pi(N_p-1)}{2(2N+1)}}^{\frac{\pi(N_p-1)}{2(2N+1)}} \frac{(e^{2ik}+e^{-2ik})}{2}\nonumber\\
&=\frac{1}{N_p}\frac{\sin\left(\frac{\pi N_p}{2N+1}\right)}{\sin\left(\frac{\pi}{2N+1}\right)}  .
\label{ns_v0_odd}
\end{align}

When o-rings 
are considered, we find that at all fillings f
the structure factor $S^*(Q)$  shows a sharp increase 
at a critical value of repulsion $V_c$; concomitantly, there is a sharp drop in the superfluid fraction $n_s$ at the same critical repulsion $V_c$
(see Fig. \ref{ss_fig2}). For finite systems, at large 
 V,
 $n_s$ values are the same for e-rings and o-rings at fillings ${\rm f}=N_p/(2N)$ and ${\rm f}=N_p/(2N+1)$, respectively.

\begin{table}[t]
\begin{center}
{
\begin{tabular}{|c|c|c|c|c|}
\hline
\multicolumn {1}{|c|}{$N_p/N_s$} & {Analytical} &
\multicolumn{3}{|c|}{Numerical value of $[S(Q)]_{\rm max}$} \\ \cline{3-5} 
{} & {value of $[S(Q)]_{\rm max}$} &{ V=50} & { V=100} & {V=500}\\
\hline
$\frac{6}{15}$ & $4.4828$ & $4.4574$ & $4.4618$ & $4.4650$ \\
\hline
$\frac{6}{17}$ & $3.8866$  & $3.8634$ & $3.8691$ & $3.8734$  \\
\hline
$\frac{6}{19}$ & $3.4302$  & $3.4124$ & $3.4183$ & $3.4228$  \\
\hline
$\frac{6}{21}$ & $3.0697$  & $3.0566$ & $3.0623$ & $3.0668$  \\
\hline\hline
$\frac{8}{19}$ & $6.0982$  & $6.0702$ & $6.0740$ & $6.0766$  \\
\hline
$\frac{8}{21}$ & $5.4572$  & $5.4264$ & $5.4316$ & $5.4355$  \\
\hline
\end{tabular}}
\caption{{At filling fraction ${\rm f}=\frac{N_p}{(2N+1)}$ with $N_p < N$,
 the analytical value of $[S(Q)]_{\rm max}$ [obtained from Eq. (\ref{sk_o})]
 approximates reasonably well the numerical value at large $V$.}}
\label{table_sf2}
\end{center}
\end{table} 

In the thermodynamic limit,
we expect  a first order CDW transition, with the structure factor jumping at a critical $V$
 similar to the case in Fig. \ref{ss_fig} for  e-rings
 (although, magnitude-wise, $[S(Q)]_{\rm max}$ for o-rings
is $\frac{4}{\pi^2}$  of 
the structure factor $[S(\pi)]_{\rm max}$ for e-rings);
simultaneously, at the same critical $V$ we expect a sudden 
drop in $n_s$ similar to Fig. \ref{ss_fig}.
At filling $N/(2N+1)$ the superfluid fraction 
 decreases to zero, whereas at all lower fillings [i.e., $N_p/(2N+1)$ with $N_p < N$] it 
transits to a nonzero value.
Thus,
 at filling $N/(2N+1)$, superfluidity and CDW state are mutually exclusive; whereas, at all fillings $N_p/(2N+1)$ with $N_p < N$
 the system 
undergoes a  transition from a superfluid to a supersolid.


\begin{figure}[b]
\includegraphics[width=0.7\linewidth,angle=-90]{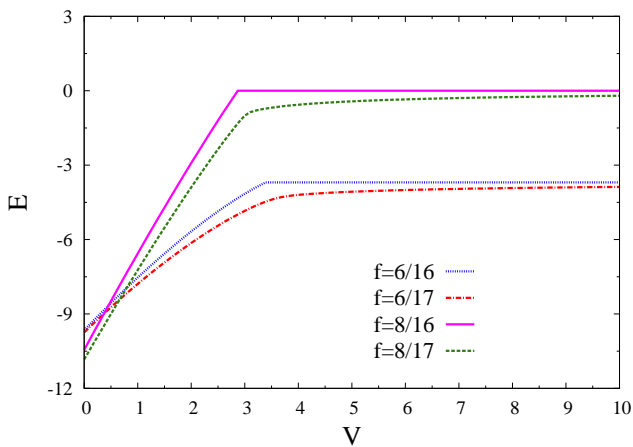}
\caption{(Color online) Plots of  energy E versus repulsion V for groundstate at various fillings ${\rm f}=N_p/N_s$.} 
\label{ev_fig}
\end{figure}


\section{Comparison between \lowercase{e-rings} and \lowercase{o-rings}}\label{sec6}
In our $t_2-V$ model, e-rings have  two similar homogeneous Fermi seas whose occupational symmetry is broken
at a critical repulsion to populate only one non-interacting Fermi sea thereby
producing a supersolid (CDW) at non-half (half) filling. The situation for o-rings,
corresponds to a single band breaking up into two bands with a midgap state.

From the energy versus V plots for e-rings (in Figs. \ref{e-rings_fig} and \ref{e-rings_fig2}), 
we note that for different finite systems with even number of sites (both for half and non-half fillings) 
the various lowest energy levels [with particle distribution ($m,N_p-m$)] cross at a critical $V$;
from this and our Green's function analysis above, we conclude that in the thermodynamic 
limit the system will undergo a first order phase transition at all fillings.
On the other hand, for o-rings, since the system has all sites
connected through NNN hopping (much
like a Moebius strip), there do not exist two sub-lattices.
Consequently, when we plotted  the lowest few energy states for a finite  odd number of sites,
there is no energy level crossing at any repulsion V for various fillings considered; the curves monotonically increase with repulsion [see Fig. \ref{o-rings_fig} and \ref{o-rings_fig2}].

For fillings ${\rm f}=N/(2N+1)$,
 we observe that beyond some critical repulsion (close to the value where levels cross in  half-filled e-ring with $N_s=2N$)
all the energy levels merge and become degenerate. For filling 
fractions  ${\rm f}=N_p/(2N+1)$ (with $N_p < N$), the energy levels come close to each other at a critical V
that is close to the point where levels cross for  e-rings with ${\rm f}=N_p/(2N)$; 
a little beyond this critical V, the 
gap between any two energy levels becomes constant and remains the same even at large V. The insets in 
Figs. \ref{o-rings_fig}(a),  \ref{o-rings_fig}(b), \ref{o-rings_fig2}(a), and \ref{o-rings_fig2}(b),
show that 
at large V  the graphs 
remain more or less unchanged. Thus, for finite systems with odd number of sites, the energy levels never cross each other. 
Still, we expect that in the thermodynamic limit the ground state of an o-ring will
be similar to that of an e-ring and thus will have a kink at the same critical repulsion $V_c$.
As a result, in the thermodynamic limit, we expect both the systems to have a similar kind of phase transition. 
Another important point to note, as $V\to\infty$, is that 
the ground state energy of o-rings at ${\rm f}=N_p/(2N+1)$ approaches (from below) the ground state energy of  e-rings at ${\rm f}=N_p/(2N)$
[see Fig. \ref{ev_fig}].
 Consequently, the superfluid fraction takes the same value 
for both the cases in the limit of large V (see Table \ref{table_sup}).


\begin{table}
\begin{center}
{
\begin{tabular}{|c|c|c|c|}
\hline

\multicolumn {1}{|c|}{$N_p/N_s$}&
\multicolumn{3}{|c|}{Superfluid fraction $n_s$} \\ \cline{2-4} 
{} & { V=50} & { V=100} & { V=500}\\
\hline
$\frac{6}{16}$ &  $0.3080 $ & $0.3080 $ & $0.3080$  \\ \hline
 $\frac{6}{17}$ &  $ 0.3114$ & $ 0.3096$ & $0.3080 $  \\ \hline\hline
$\frac{6}{18}$ &  $0.4220 $ & $0.4220 $ & $0.4220 $ \\ \hline
 $\frac{6}{19}$ &  $0.4257 $ & $ 0.4240$ & $ 0.4226$  \\ \hline\hline
 $\frac{4}{16}$ &  $ 0.6533$ & $ 0.6533$ & $0.6533 $  \\ \hline
 $\frac{4}{17}$ &  $0.6574 $ & $0.6554 $ & $ 0.6537$ \\ \hline\hline
$\frac{4}{20}$ &  $0.7694 $ & $ 0.7694 $ & $0.7694  $ \\ \hline
 $\frac{4}{21}$ &  $ 0.7721$ & $ 0.7708$ & $0.7698 $  \\ \hline
\end{tabular}}
\caption{{The superfluid fraction $n_s$ for o-ring at filling ${\rm f}=N_p/(2N+1)$
 approaches the value of the superfluid fraction for e-ring with ${\rm f}=N_p/(2N)$ in the large-$V$ limit.}}
 \label{table_sup}
\end{center}
\end{table}
However, in the macroscopic limit, the energy spectrum is not expected to be the same for the two cases; for instance, there is a mid-gap state for the odd case which is not there for the even case . The key difference between the rings with even and odd number of sites  seems to be the difference in the peak value of the structure factor at large $V$ (above critical $V$) for finite (infinite) systems.
In the thermodynamic limit we expect the structure factors for both o-rings and e-rings to diverge, but their ratio  
will still be $\frac{4}{\pi^2}$.

For finite e-rings, it is also important to point out our finding that the energy levels cross each other
at approximately the same critical repulsion. From the plots of E versus V 
 at a fixed filling fraction f and  
different system sizes $N_s$ [such as in Figs. \ref{e-rings_fig}(b) and \ref{e-rings_fig2}(b)],
 we observe that the crossing points get closer as the system size increases. 
From this we can expect that, in the thermodynamic limit, all the energy levels will cross at the same transition point. 


\section{Bose-Einstein condensation}\label{sec7}
Here, we do not calculate the Bose-Einstein condensate occupation 
number $n_0$ because, for a system of HCBs in a one-dimensional tight-binding lattice, it varies as $C({\rm f}) \sqrt{N}$ in the
thermodynamic limit with the coefficient $C({\rm f})$ depending on filling ${\rm f}$ \cite{lenard,muramatsu1}; 
consequently, the condensate fraction $n_0/N_p \propto 1/\sqrt{N} \rightarrow 0$.
 Next, in the presence of repulsion (as argued in Ref. \onlinecite{srsypbl2}), we expect  the BEC occupation number $n_0$
to again scale as  $\sqrt{N}$; however, the coefficient of $\sqrt{N}$ will
be smaller due to the restriction on hopping imposed by repulsion.

\section{Connection to other models}\label{sec8}
\begin{figure}
\includegraphics[width=0.75\linewidth,angle=-90]{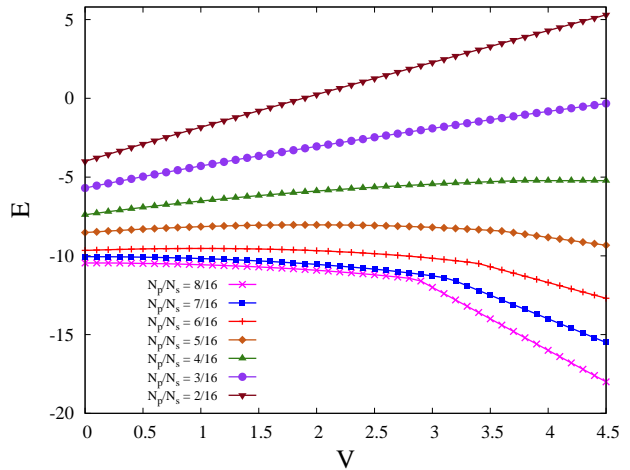}
\caption{(Color online) Plots of the lowest energy ${\rm E}$ (obtained using modified Lanczos in a system with $N_s=16$ sites) for the extremely 
anisotropic NNN Heisenberg model
at various normalized magnetizations $(N_s-2N_p)/N_s$ corresponding to fillings ${\rm f} = N_p/N_s$ for the $t_2-V$ model.
}
\label{eng_fig}
\end{figure}

Our $t_2-V$ model can be mapped onto an extremely anisotropic Heisenberg
model (with next-nearest-neighbor XY interaction and nearest-neighbor Ising interaction)
by identifying
$S^{+}=b^{\dagger}$, $S^{-}= b$, and $S^z= n - \frac{1}{2}$.
The resulting spin Hamiltonian is of the form
\be
- t_2 \sum_{i=1}^{N_s} (S^+_{i-1} S^-_{i+1} + {\rm H.c.}) + V \sum_{i=1}^{N_s} S^z_i S^z_{i+1} .
\ee
While Heisenberg model was amenable to solution through the Bethe ansatz, the addition of 
next-nearest-neighbor interaction (similar to the case of Majumdar-Ghosh model \cite{ckm})
requires an alternate route for its solution. 
Our spin model 
 lends itself to exact instability solutions (by the Green's function method)
in the two limiting cases of two-magnons and antiferromagnetic ground state. 
The energies at various fillings $N_p/N_s$ for the $t_2-V$ model correspond to various 
normalized magnetizations $(N_s-2N_p)/N_s$ for the spin model. From a plot of the lowest energies at various magnetizations
of the spin model, as depicted in Fig. \ref{eng_fig},  we see that the energy increases as the normalized magnetization increases
with the ground state corresponding to zero magnetization.  From the fact that
 the critical repulsion is always $V_c \le 4$, it should be clear that the energy at all fillings
and system sizes for $V_c > 4$ is obtained from a tight-binding model with $N_p$ particles in one sub-lattice only.
Thus at $V_c > 4$, the energy will certainly increase with the normalized magnetization.


Next, in spite of the fact that the hopping terms are different,
we note the semblance between our $t_2-V$ model and 
the well-known Su-Schrieffer-Heeger (SSH) model \cite{ssh}.
We make the connection
that a singlet (formed by two spins on adjacent sites) can be regarded as a HCB located at the
center of the singlet \cite{srsypbl2}. Thus a system of HCBs in one sublattice
is transformed to a system of singlets with centers located in one sublattice only.
At half-filling, for even number of sites, the ground state of the $t_2-V$ model at $V$ larger than the critical repulsion
can be mapped onto a valence-bond ground state of the Su-Schrieffer-Heeger (SSH) model \cite{ssh}.
 For an odd number of sites, at filling $N/(2N+1)$, we can map the ground state
at large $V$
with  two holes on adjacent sites (i.e., a kink) in the $t_2-V$ model
on to the ground state, with a kink or  topological defect (with two single bonds on adjacent sites), in the SSH model.
For even number of sites as well, the kinks obtained by doping the valence bond state in the SSH model have a counterpart in our $t_2-V$ model. 


\section{Discussion}\label{sec9}

{Our $t_2-V$ model is the limiting case of the one-dimensional
 CBM (cooperative breathing mode) model
(i.e., $t_1-t_2-V$ model), which depicts a simpler one-band case that is expected to be useful in 
understanding the CBM physics in real systems such as the bismuthates, the cuprates, and 
the manganites \cite{rpsy}. An important purpose of studying the $t_2-V$ model is the 
fact that exact solutions can be obtained analytically
for two limiting cases of this model.
 Till now we do not know how to solve the more complex $t_1-t_2-V$
model analytically. We hope that in future our way of solving the two limiting cases of the $t_2-V$ model
 analytically
(using Green's function technique) will lead to useful
approaches to handle more complex problems such as the $t_1-t_2-V$ model.

In Ref. \onlinecite{rpsy}, $t_2-V$ model was studied for spinless fermions in e-rings; 
structure factor $S(\pi)$ and  ground state energy were obtained for some filling fractions
to show that the system undergoes a discontinuous transition from a Luttinger liquid to a
conducting commensurate CDW state away from half filling while at half filling one obtains a Mott insulator.
In the present paper, we recast the $t_2-V$ model in terms of HCBs. Here, along with the 
structure factor and the ground state energy, we additionally calculate the superfluid fraction.
In e-rings, we show that the system undergoes a striking discontinuous transition from a
superfluid to either a  CDW insulator (at half filling) or a supersolid (at non-half fillings). 

In the previous paper \cite{rpsy}, only systems with even number of sites was considered, 
whereas here we also study 
systems with odd number of sites; we show that supersolidity is realized in o-rings
at large repulsion and fillings $N_p/(2N+1)$ with $N_p < N$. 
When o-rings are considered, at all fillings $N_p/(2N+1)$ with $N_p < N$,
 the structure factor shows a sharp increase at a critical value of repulsion;
simultaneously, there is a sharp drop (to a non-zero value) in the superfluid fraction
at the same critical repulsion (see Fig. \ref{ss_fig2}). At $V=0$, we derived an expression
for the superfluid fraction [see Eq. (\ref{ns_v0_odd})] which matches with 
the numerical result in Fig. \ref{ss_fig2}. 
The superfluid fraction, in the limit $V \rightarrow \infty$, for o-rings [at fillings $N_p/(2N+1)$]
approaches the value of the superfluid fraction, when $V > V_c$,  for e-rings [at fillings $N_p/2N$] 
-- a fact supported by numerical results (see Table \ref{table_sup}). From the expression for the superfluid fraction in 
e-rings when $V > V_c$ [see Eq. (\ref{ns1})], it is clear that, for all fillings $N_p/(2N+1)$ with $N_p < N$,
we will have non-zero superfluid fraction in o-rings
even in the thermodynamic limit. On the other hand, we have also
shown that the structure factor $[S(Q)]_{\rm max}$ calculated analytically for large $V$ agrees
quite well with that calculated numerically for sufficiently large systems (see Table \ref{table_sf2}). It is also shown
that in the thermodynamic limit, at large $V$, the structure factor for o-rings appears to
be $4/\pi^2$ times the structure factor for e-rings [see Eq. (\ref{sk_inf})]; as a result we expect the structure factor
for o-rings to diverge as $N \rightarrow \infty$. Hence, we can conclude that in the
thermodynamic limit, at all  all fillings $N_p/(2N+1)$ with $N_p < N$,
the system for o-rings exhibits supersolidity (at large $V$).

As regards relevant work, in Ref. \onlinecite{zigzag} the authors consider nearest-neighbor 
hopping (and repulsion) and unfrustrated next-nearest-neighbor hopping (and repulsion) that 
can be realized in a zigzag ladder with two legs. As pointed out in this paper,
nearest-neighbor hopping and next-nearest-neighbor hopping can be tuned independently.
Thus this work also shows that our $t_2-V$ model is physically realizable. 

Also of relevance is the  work by Struck et al.\cite{science} where various 
values (including sign change) of nearest-neighbor coupling $J$ and next-nearest-neighbor coupling $J^\prime$
can be achieved for hard-core-boson systems. Thus, we feel that our 
$t_2-V$ model can be simulated experimentally by introducing repulsions and next-nearest-neighbor 
coupling $J^\prime$.

Recently, Mishra et al. \cite{mishra1,mishra2} studied a one-dimensional system of HCBs, 
with nearest-neighbor hopping
 and interaction and next-nearest-neighbor hopping, described by the Hamiltonian
\be
H=&-t_1\sum\limits_i(b_i^\dagger b_{i+1}+{\rm H.c.})-t_2\sum\limits_i(b_i^\dagger b_{i+2}+{\rm H.c.})
\nonumber
\\
&+\sum\limits_i V\left(n_i-\frac{1}{2}\right)\left(n_{i+1}-\frac{1}{2}\right) .
\ee
 While Pankaj and Yarlagadda \cite{rpsy} considered unfrustrated
next-nearest-neighbor hopping, Misra et al. \cite{mishra1,mishra2} focussed  on frustrated hopping.
 In Ref. \onlinecite{mishra1},
at half filling the authors set $t_1 = 1$ and varied $t_2$ from $0$ to $-t_1$ to obtain the 
total phase diagram containing superfluid, CDW, and bond-ordered phases. At incommensurate densities
(non-half fillings) and with $t_2=-t_1$, the authors of Ref. \onlinecite{mishra2}
 found a supersolid phase.
 In Refs. \onlinecite{mishra1,mishra2}, the competition between two different
hopping processes (i.e., hoppings from one site to nearest-neighbor site and to next-nearest-neighbor
site with different signs of the hopping terms) gives rise to kinetic frustration in the system.
On the other hand, our $t_2-V$ model depicts the strong EPI limit; as a result there is 
only one kind of hopping process. Consequently, there is no frustration in our $t_2-V$ system.

\section{Conclusions}\label{sec10}
We investigated a  model which captures
the essential dominant-transport feature of cooperative strong EPI in all dimensions
 (and in even non-cubic geometries). 
Our study shows that, compared to the non-cooperative situation, cooperative EPI produces strikingly
different physics such as a dramatic superfluid to a supersolid transition
with the order parameter jumping to its maximum value.
Understanding  one-dimensional strong EPI systems, besides being helpful in
designing oxide rings, will be of relevance to predicting
and controlling the variation of system properties in the direction normal to
the interface in oxide heterostructures; needless to say, oxide rings and oxide heterostructures offer extraordinary
scientific and technological opportunities \cite{millis}.
}


\section{Acknowledgments}
S.Y. acknowledges vital discussions with M. Berciu, R. Pankaj, Diptiman Sen, 
P. B. Littlewood, and G. Baskaran. S.Y.  also thanks KITP for hospitality.
This work is dedicated to the interactions with and 
 memory of Gabriele F. Giuliani.

\appendix

\section{Finite size scaling analysis for HCBs in $t_2-V$ model for \lowercase{e-rings}}
In this section, we will outline our approach to carrying out finite size scaling analysis for a system with even number of sites.
{Let us first consider a non-interacting system  (with {$2N$} sites and $N_p$ particles)
 described by a tight-binding Hamiltonian. For even number of particles,
the ground state has particles occupying momenta $ \frac{(2m+1)\pi}{2N}$ with $-N_p/2 \le m \le N_p/2 -1$; whereas
for odd number of particles the ground state is represented by particle momenta $ \frac{(2m)\pi}{2N}$
with $-(N_p-1)/2 \le m \le (N_p-1)/2$. Thus the ground state wavefunction (due to the occupied momenta)
is an even function of {$1/2N$}. Now, for the case corresponding to after the
phase transition where all the particles are in the same sublattice C, the energy
of the ground state $|\phi_0 \rangle$
is given by
\begin{eqnarray}
E_{I} = \sum_{i=1}^{N}  \langle \phi_0 | -t_2(c^{\dagger}_{i} c_{i+1} + {\rm H.c.}) |\phi_0 \rangle  ,
\end{eqnarray}
where $c$ is the destruction operator for a HCB in sub-lattice C.
Upon taking into account discrete translational symmetry, we get
\begin{eqnarray}
\frac{ E_{I}}{N} =  \langle \phi_0 | -t_2(c^{\dagger}_{i} c_{i+1} + {\rm H.c.}) |\phi_0 \rangle . 
\end{eqnarray}
Since $|\phi_0 \rangle$ is even in {$1/2N$}, we note that $\frac{E_{I}}{N}$ is also even in {$1/2N$}.
 
Next, consider the interacting system characterized by the
following $t_2-V$ Hamiltonian
in rings with even number of sites ({$2N$}) and with periodic boundary conditions:
\begin{eqnarray}
H \equiv &&  
 - t_2 \sum_{i=1}^{N} (c^{\dagger}_{i} c_{i+1} + {\rm H.c.} ) 
- t_2 \sum_{i=1}^{N} (d^{\dagger}_{i}d_{i+1}  + {\rm H.c.} ) 
\nonumber \\
&&
+V \sum_{i=1}^{N} d^{\dagger}_i d_{i} (c^{\dagger}_i c_i + c^{\dagger}_{i-1} c_{i-1}) ,
\label{eq:tVapp}
 \end{eqnarray}
where $c$ ($d$) is the destruction operator for HCB in sublattice $C$ ($D$) and $V \ge 0$.
Upon invoking reflection symmetry, we note that the ground state will be invariant when the sign of momenta
is reversed; equivalently the ground state $|\psi_0 \rangle$ is an even function of {$1/2N$}.
The ground state energy, before the phase transition, is given by
\begin{eqnarray}
 E_{II} &=& \sum_{i=1}^{N}  \langle \psi_0 | -t_2[(c^{\dagger}_{i} c_{i+1} + {\rm H.c.}) + (d^{\dagger}_{i}d_{i+1}  + {\rm H.c.} )]|\psi_0 \rangle 
\nonumber \\
 && ~+ \sum_{i=1}^{N} \langle \psi_0 | V d^{\dagger}_i d_{i} (c^{\dagger}_i c_i + c^{\dagger}_{i-1} c_{i-1}) | \psi_0 \rangle .
\end{eqnarray}
Upon recognizing discrete translational invariance, we see that
\begin{eqnarray}
 \frac{E_{II}}{N} &=&   \langle \psi_0 | -t_2[(c^{\dagger}_{i} c_{i+1} + {\rm H.c.}) + (d^{\dagger}_{i}d_{i+1}  + {\rm H.c.} )]|\psi_0 \rangle 
\nonumber \\
 &&~+  \langle \psi_0 | V d^{\dagger}_i d_{i} (c^{\dagger}_i c_i + c^{\dagger}_{i-1} c_{i-1}) | \psi_0 \rangle .
\end{eqnarray}
Since, $|\psi_0\rangle$ is even in $1/2N$, it follows that $\frac{E_{II}}{N}$ is also even in $1/2N$. 

Now, at the transition point (corresponding to a critical interaction $V_c$), $\frac{E_{II}}{N}-\frac{E_{I}}{N} =0 $; this implies that
\begin{eqnarray}
 V_c &=& \frac{\langle \psi_0 | t_2[(c^{\dagger}_{i} c_{i+1} + {\rm H.c.}) + (d^{\dagger}_{i}d_{i+1}  + {\rm H.c.} )]|\psi_0 \rangle} 
{\langle \psi_0 |  d^{\dagger}_i d_{i} (c^{\dagger}_i c_i + c^{\dagger}_{i-1} c_{i-1}) | \psi_0 \rangle}
\nonumber \\
&&-\frac{\langle \phi_0 | t_2(c^{\dagger}_{i} c_{i+1} + {\rm H.c.}) |\phi_0 \rangle}
{\langle \psi_0 |  d^{\dagger}_i d_{i} (c^{\dagger}_i c_i + c^{\dagger}_{i-1} c_{i-1}) | \psi_0 \rangle} .
\end{eqnarray}
In the above equation, because both numerator and denominator of all the terms on the right-hand side
are even in $1/2N$, it follows that $V_c$ is
also even in $1/2N$.
Thus, for e-rings at various fillings,
we relate $V_c(2N)$ (critical repulsion at $N_s=2N$) to  $V_c(\infty)$ as follows: 
\be
 V_c(2N) -V_c (\infty) = \frac{A}{N^2} + \frac{B}{N^4} + .....  ,
\label{Vc}
\ee
where $A$, $B$,... are constants.

It is also important to note that we used general arguments to show that 
the ground state energy $\frac{E_{II}}{N}$ is  even in $1/2N$;
these arguments can be extended to show that the ground state energies of other interacting systems such as the $t-V$ model
are also even functions of $1/2N$. The fact that the ground state energy of the $t-V$ model is an even function
of $1/2N$ has been used in Ref. \onlinecite{dagotto}.


\end{document}